# Statistical Image Analysis of Drying Bovine Serum Albumin Droplets in Phosphate Buffered Saline

Anusuya Pal[a], Amalesh Gope[b*], and Germano S. Iannacchione[a]

[a] Order-Disorder Phenomena Laboratory, Department of Physics,

Worcester Polytechnic Institute, Worcester, MA, 01609, USA

[b] Tezpur University, Tezpur, Assam, 784028, India

Abstract

A bio-colloidal drying droplet can be used as a pre-diagnostic technique. However, a successful clinical setting requires a fundamental understanding of the final morphology and the way it is related to the initial state of the constituents present in the droplet. This chapter focuses on the physics associated with different pattern formations in the globular protein, bovine serum albumin (BSA) at different phosphate buffered saline concentrations. The study reports that the first-order statistics (FOS) and the gray level co-occurrence matrix (GLCM) analysis are capable of capturing structural changes of the droplets. While the FOS of the image depend on the individual pixels, the GLCM summarizes both tonal and structural relationships between the neighboring pixels. The horizontal and the vertical orientations of the GLCM parameters show a non-significant effect when the pixel displacement is $\leq 1$. Interestingly, two local equilibrium-like regions (the rim and the central regions) appear when these droplets approach the steady state. The bimodal distribution confirms that the BSA-BSA interactions are dominant (recessive) over the BSA-saline interactions in the rim (central) regions that result in the phase separated aggregation of the BSA particles in the presence of the salts.

*Corresponding author: email id- amaleshgope5sept@gmail.com

## I. INTRODUCTION

Any droplet, ranging from a droplet of spilled coffee, or orange juice to the bloodstain on the crime spots, are a few examples of colloids in the sessile drop configuration [1,2]. The evaporative drying of the solvent in such droplets may appear simple; however, the physical mechanisms involved in the drying mechanism is hard to understand [3]. The complication arises due to the coupling of the droplet with the substrate on which it sits- the environmental conditions (temperature, atmospheric pressure, relative humidity, etc.), and the constituent particles present in the droplet led to the complex mechanism [4]. During the drying process, the colloidal droplet gets pinned to the substrate in such a way that the droplet radius can't shrink [5]. In this process the solvent evaporates, and the particles present in the droplet distribute themselves. The drying mechanism shows distinct patterns depending on the nature of the constituents. Nonetheless, the mechanisms involved in such drying colloidal droplets cannot be generalized due to their large-scale variability. In most of the cases, the constituent particles arrange themselves in such a way that the droplets show a well-known "coffee-ring" [6] behavior. In some cases, these particles configure as a film, and cover the whole droplet. Furthermore, some droplets experience high mechanical stress and crack along their weak failure lines; whereas, some do not. The underlying mechanism in such droplets is modeled analytically and numerically by several researchers [7–12]. Nevertheless, it becomes a challenging task when these colloids self-interact [13-14]. This self-assembly is typical in most of the bio-colloids, such as protein, serum, plasma, blood, etc. A detailed investigation, therefore, is the need of the hour to explore the final morphology of these bio-colloidal droplets (under a given set of external conditions such as temperature, relative humidity, substrate type, etc.) is related to the initial state of the constituent particles.

Recent advances in the bio-colloidal drying droplets have shown that these morphological patterns have the potential to predict an individual's risk of a pathological condition (cancer, pneumonia, glaucoma, etc.) or even identifying a crime scene by tracking the evolving patterns within the droplet [15–23]. All these pieces of evidence support the possible application of this methodology to be used as a pre-diagnostic or screening technique in the near future. However, a successful adaptation of a clinical setting depends on the investigation of a large-sized sample data. The drying evolution and the final morphology need to be identical for each set of given conditions. To this end, some researchers have focused their work on the fundamental understanding of the patterns evolved in the simple bio-colloids (aqueous solutions of the proteins or the proteins in the ionic solutions) [24–33]. In contrast, a few other researchers have investigated human blood plasma and blood; and explored the distinct patterns developed due to the change in the relative humidity, the wetting, the ionic concentration, etc. [34–41].

The standard assessment protocols of both the drying evolution and the resulting morphological patterns are indicated through mass loss, contact angle, contact line, fluid front progression, etc. Until very recently, the drying mechanism of the protein samples is primarily explained with the help of two globular water-soluble proteins; viz., [lysozyme and bovine serum albumin (BSA)] in presence or absence of different types of salts [31, 33, 42, 43]. Yakhno et al. [28] reported the presence of a 'phase transition' while investigating the distribution of BSA and NaCl salts in the drying droplet and concluded that the proteins are phase-separated from these salts. The BSA particles are mostly observed near the periphery, while the crystal-like structures are formed in the central region of the droplet. Chen et al. [29] examined BSA proteins in the presence of different phosphate buffer saline (PBS) and observed two distinct drying modes during the interaction phase with the salt residues. The quicker drying rate is detected near the edge of the droplets when the initial salt concentration is low. This evaporation rate is greater at the center of the droplets at higher salt concentrations. In a recent study, Carreón et al. [32] examined the mixture of BSA and lysozyme proteins in presence of NaCl salts using a new textural image analysis. The study reports that the first-order statistics (FOS) and the gray level co-occurrence matrix (GLCM) analysis are capable of capturing structural changes due to the formation of complex dendrite, rosette, and scalloped-like structures in such drying droplets.

Despite intense research on the protein drying droplets, only a few studies exploited the image processing techniques to explore the drying mechanism [44-46]. The images of any droplets captured at different drying stages are potential pattern recognizing tools, nonetheless, the data extraction procedure of those images is very complex and sensitive task. For example, the GLCM of an image is highly dependent on its orientation and pixel displacement [47-48]. To the best of our knowledge, no study till date attempted to examine these (initialized) factors of the GLCM parameters. Furthermore, the gray values in terms of the pixel counts are not explored for the drying droplets.

To elucidate this gap, this chapter focuses on the physics associated with different pattern formation by exploiting the FOS and GLCM textural image parameters. Our principal motivation is to explore the effects of different saline buffer concentrations (Ø) from 1x to 5x on the time evolution and the morphological properties of BSA protein during the drying process using optical bright-field microscopy. Additionally, we also wish to develop a mechanism that can interpret the pixel distribution of the images acquired during the drying process. Our final goal is to examine the influence of horizontal (0°) and vertical (90°) orientations and pixel displacements (1, 10, 50, and 1000) on such GLCM parameters using suitable statistical tests. The FOS parameters (viz., the mean (M), standard deviation (SD), kurtosis (KUR), and skewness (SKEW)), and the GLCM parameters (viz., angular second moment (ASM), correlation (COR), inverse difference moment (IDM), and entropy (ENT)) are thoroughly examined in this chapter.

Following this introduction, Section II describes the experimental methodology adopted in this chapter. The qualitative and quantitative results during the drying evolution and the dried states are presented in Section III. The underlying physical mechanism is discussed in Section IV, and the conclusions are drawn in Section V.

## II. EXPERIMENTAL METHODS

The bovine serum albumin (BSA) is one of the well-explored blood globular protein. Each BSA particle is a prolate ellipsoid that carries a net negative charge (pH of 7.3-7.5) at the time of experiments [11]. 10x PBS contains a mixture of different salts, i.e., 0.137 M NaCl, 0.0027 M KCl, and 0.119 M phosphates (purchased from the Fisher BioReagents, USA; Catalog number BP243820). BSA was bought from the Sigma Aldrich, USA (Catalog number A2153). 10x PBS was diluted to different saline concentrations (Ø) of 1x, 2x, and 5x. The protein, weighed ~100 mg, was mixed with different Ø. The prepared samples were used within a few hours. A volume of ~1 μL of the sample was pipetted on a clean coverslip (Erie Scientific, USA, Catalog number 48366-045) to form a ~1 mm radius droplet under ambient conditions (room temperature of 25°C and relative humidity of 50%). After its deposition, the time-lapse images were recorded at every two seconds under the bright field microscopy (Leitz Wetzlar, Germany). The microscopy operates in the transmission mode with a 5x objective lens. The images of the dried films were captured after 24 hours of the drying process. The experiments were repeated twice to ensure even reproducibility. These images were captured using an 8-bit digital camera (Amscope MU300) at a fixed resolution of 3664 x 2748 pixels. A region of interest was determined on the image using the oval tool in ImageJ [49]. The histogram depicting the pixel counts of the gray values is obtained from 8-bit, gray-scaled time-lapse images at different drying times. The gray values of 8-bit images ranges from 0 to 255. The FOS parameters [the mean (M), standard deviation (SD), kurtosis (KUR), and skewness (SKEW)] are also extracted from these 8-bit gray images. All the GLCM parameters [the angular second moment (ASM), correlation (COR), inverse difference moment (IDM), and entropy (ENT)] are computed at various pixel displacements (1,10, 50, 100, and 1000) along the horizontal ($0°$) and vertical ($90°$) orientations using Texture Analyzer plugin in ImageJ.

The non-parametric Mann Whitney U statistical test is performed to see the significant differences (if any) between these orientations when the displacements are varied at each Ø. The width of the ring (w) and the radius of the droplet (R) were measured five times, and their averaged values were computed. The spacing between the consecutive radial cracks in the ring was calculated using ImageJ. The detailed procedure can be found in [44-46]. The normalized rim width ($\overline{w}/\overline{R}$) and the averaged crack spacing ($\overline{x}_c$) were plotted as a function of Ø.

## III. RESULTS

### A. Temporal study of the drying droplets

Figure 1(I-III) shows the pixel counts as a function of gray values for BSA-saline droplets at different initial PBS concentrations (Ø of 1x, 2x, and 5x). The images of the time evolution during the drying process are displayed in the insets of Fig.1(I-III) a-e. The first image was captured within ~1.5 minutes after the deposition of these droplets on the substrate (coverslip). The droplets show dark gray shade near the periphery. The light gray texture covers most of the droplet's area at each Ø (Fig. 1(I-III) a). The histograms [black in (I), yellow in (II), violet in (III)]) extracted from the images (Fig. 1(I-III) a) indicate a gaussian distribution between ~90 and ~130 gray values with a flatter tail on the left side of the distribution. The fluid front recedes from the periphery towards the center within ~1 minute of the droplets' deposition. The movement continues for ~3 minutes. The white dashed curved lines in Fig. 1(I-III) b mark its movement from the periphery to the center. The distribution in the tail is reduced. Moreover, the symmetry in the distribution disappears when the histogram is compared between Fig. 1(I-III) a and b. As time progresses, more water evaporates from these droplets. A bulged ring near the periphery is observed. Two different regions (the rim and the center) are established. These regions are marked with a dashed yellow circular line. Furthermore, the equally spaced radial cracks and a few orthoradial cracks near the periphery are noticed (Fig. 1(I-III) c). The fluid front continues to move while the radial cracks propagate, and intervene through the rim to the central region at each Ø. Interestingly, a spike at a gray value of ~200 is observed (Fig. 1(I-III) c). The grainy structures start forming near the rim's inner edge, and these structures cover the entire central region (Fig. 1(I-III) d). As the salt crystals grow more and more, the texture becomes dark, and the pixel values decrease. The pixel counts also reduce when the saline concentration (Ø) increases from 1x to 5x. The final morphology is presented at the end of the visible drying process (Fig. 1(I-III) e). The texture of the rim is transparent at Ø = 1x (Fig. 1(I) e) and does not contain any optically visible structures like what observed in the BSA film prepared with only de-ionized water [11]. This texture becomes less transparent with the increase of the saline concentration from 2x to 5x (Fig. 1(II-III) e). The central region appears to be grainy near the dashed yellow line and (mostly) gray in the center at Ø = 1x. In contrast, this region turns out to be inhomogeneous, and the texture becomes dark gray with the upsurge of Ø. The radial cracks are observed to propagate towards the central region at Ø of 1x and 2x; however, it is hard to conclude anything for 5x (Fig. 1(III) e). The pixel distribution changes to gaussian at 1x and 2x (Fig. 1(I-II) e). Surprisingly, a bi-modal distribution is observed at 5x, where the first and second peaks are observed at ~20 and ~120 gray values, respectively (Fig. 1(III) e).

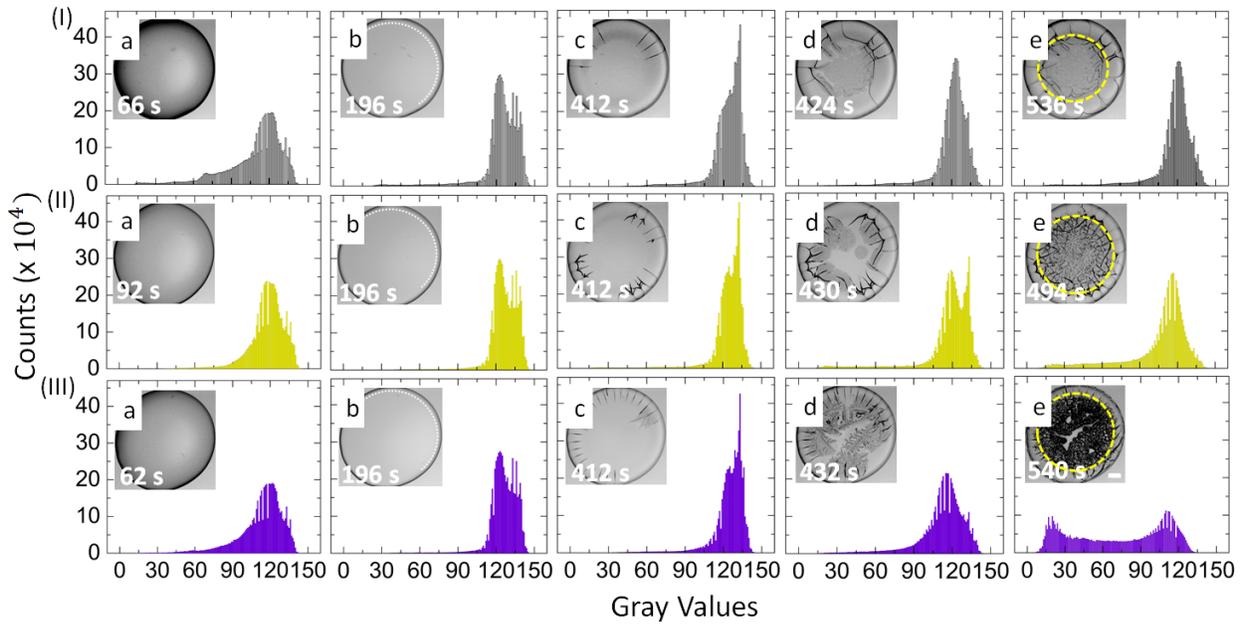

**Figure 1.** BSA saline droplets at different initial PBS concentrations (Ø): The insets show the drying evolution of the droplets is displayed in a-e at Ø of (I) 1x, (II) 2x, and (III) 5x. The first image is taken within ~90 seconds after their deposition on the substrate and is presented in a. The movement of the fluid front is indicated with the white dotted curved lines in b. The appearance of the radial cracks in the periphery and their inward propagations is mostly seen in c. The origin of the crystals divides the droplets into the central and the rim regions. The texture of the droplets changes due to the presence of the salts and is described in c-d. The visible drying process takes

~600 seconds. A dashed circular yellow line outlines the rim. The images are taken after completion of the process in e. The time at which the images are captured is shown at the left corner in every image. The scale bar in (III)e corresponds to 0.2 mm. The different colored histograms depict the counts of the pixels along the y-axis, and the gray values along the x-axis of each image in a-e at various Ø.

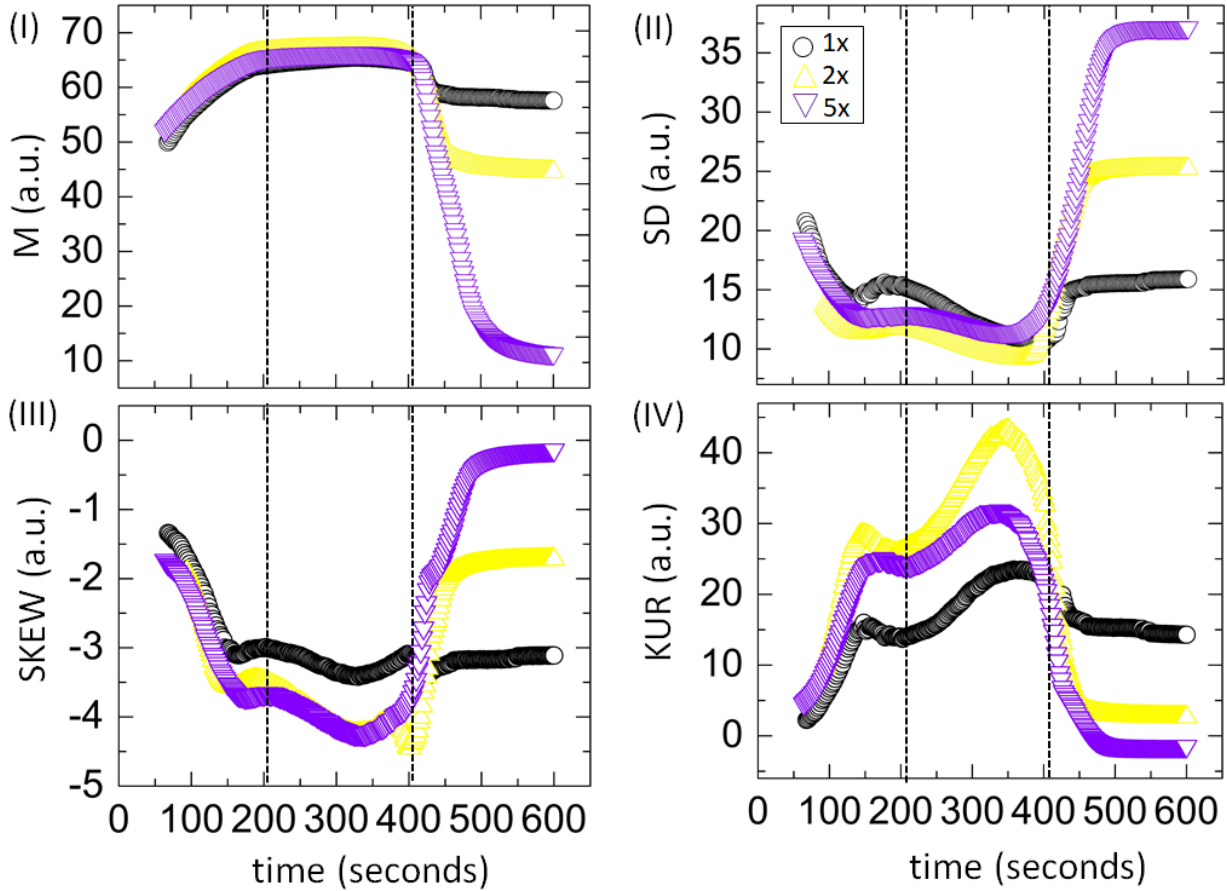

**Figure 2.** (I-IV) shows the time evolution of the first-order statistical (FOS) parameters [mean (M), standard deviation (SD), skewness (SKEW), and kurtosis (KUR)] in arbitrary units (a.u.), respectively in the BSA saline droplets as a function of drying time at the different initial saline concentration (Ø) of 1x to 5x. The M begins with a slow increase, followed by a nearly constant

value from ~200 seconds to ~400 seconds. It diminishes as the drying evolution progresses towards the end. The different stages of the drying process are outlined with the dashed lines in I-IV.

### A. FOS characterization of the drying evolution

Figure 2(I-IV) exhibits the first-order statistics (FOS) parameters (mean (M), standard deviation (SD), skewness (SKEW), and kurtosis (KUR)] respectively) as a function of drying time at the different initial saline concentration (Ø) of 1x to 5x.

The M in Fig. 2(I) increases from ~50 arbitrary units (a.u.) to ~65 a.u. till ~200 seconds. It remains the same for the next 200 seconds (~200 seconds to ~400 seconds, outlined with dashed lines). Interestingly, the M does not indicate any significant change for different Ø until ~400 seconds. After ~400 seconds, it starts decreasing from ~65 a.u. to ~50 a.u. for 1x and 2x, whereas it reduces to ~10 a.u. at Ø = 5x. The SD, on the other hand, depends on different Ø before ~400 seconds unlike the M [Fig. 2(I-II)]. The SD initially displays a decrease from ~20 a.u. to ~10 a.u., whereas an upsurge is observed in the later time from ~400 seconds [Fig. 2(II)]. The SKEW in Fig. 2(III) illustrates the negative values that decrease from -1.5 a.u. to -4.5 a.u. till ~400 seconds; however, it starts increasing afterward. On the other hand, the KUR in Fig. 2(IV) shows an increasing trend until ~350 seconds. It declines as the drying process ends. It is worthy to mention that the M shows three ubiquitous stages (illustrated with dotted lines at ~200 seconds and ~400 seconds). In contrast, the other parameters (SD, SKEW, and KUR) reveals multiple peaks and dips throughout the drying process.

### B. GLCM characterization of the drying evolution

The GLCM (Gray Level Co-occurrence Matrix) parameters are the second-order statistics calculated from the spatial relationship between two neighboring pixels [32]. This makes GLCM parameters complicated and different from the first-order statistics (FOS), which solely depends on the individual pixel values [47-48]. Before establishing any relationship between two neighboring pixels, it is crucial to address a few things. For example, what will be the displacement between the neighboring pixels? Should the displacement be close at each other? What if the displacement between the pixels is far enough? Are there any limitations? What are the directions of the pixels? Which pixels should be counted- the pixels which are placed horizontally or vertically, or their average? In our study, we examined how the directions can be used in extracting the GLCM parameters such as the angular second moment (ASM), correlation (COR), entropy (ENT), and inverse difference moment (IDM) at various initial concentrations (Ø) from 1x to 5x. Furthermore, a non-parametric Mann–Whitney U test is performed to see whether these GLCM parameters are rotationally invariant along the horizontal (0°) and the vertical (90°) orientations for the BSA-saline droplets. In the test, the rotation was kept as the independent factor, with two levels, 0°, and 90°. The GLCM parameters are chosen as the dependent variables. The mean over the median rank in the test is chosen as the number of time points during the drying process is observed to be relatively large. The p-value of less than 0.05 is considered to be significant interaction [indicated with an asterisk (*)]. The detailed report of the statistical test (Mann-Whitney U, Wilcoxon W, Z, and p-values) is shown in the Tables T1- T4.

| Concentration | Displacement (pixels) | Mann-Whitney U | Wilcoxon W | Z | Asymp. Sig. (2-tailed) p value |
|---|---|---|---|---|---|
| 1x | 1 | 33231.000 | 69546.000 | -1.636 | .102 |
| 1x | 10 | 33215.500 | 69530.500 | -1.645 | .100 |
| 1x | 50 | 31300.500 | 67615.500 | -2.707 | .007* |
| 1x | 100 | 30750.000 | 67065.000 | -3.012 | .003* |
| 1x | 1000 | 194.000 | 36509.000 | -19.961 | .000* |
| 2x | 1 | 42802.500 | 87055.500 | -.623 | .534 |
| 2x | 10 | 41722.000 | 85975.000 | -1.139 | .255 |
| 2x | 50 | 39085.000 | 83338.000 | -2.400 | .016* |
| 2x | 100 | 39288.000 | 83541.000 | -2.303 | .021* |
| 2x | 1000 | 6857.500 | 51110.500 | -17.810 | .000* |
| 5x | 1 | 43900.000 | 88153.000 | -.098 | .922 |
| 5x | 10 | 43633.000 | 87886.000 | -.225 | .822 |
| 5x | 50 | 40752.000 | 85005.000 | -1.603 | .109 |
| 5x | 100 | 40508.000 | 84761.000 | -1.720 | .085 |
| 5x | 1000 | 15612.000 | 59865.000 | -13.624 | .000* |

Table: 1 Detailed report of Mann Whitney U test for ASM (angular second moment) at each concentration (1x to 5x) and displacement from 1 to 1000 (in pixels). The significant values are marked with a red colored asterisk.

| Concentration | Displacement (pixels) | Mann-Whitney U | Wilcoxon W | Z | Asymp. Sig. (2-tailed) p value |
|---|---|---|---|---|---|
| 1x | 1 | 33647.000 | 69962.000 | -1.405 | .160 |
| 1x | 10 | 22395.500 | 58710.500 | -7.646 | .000* |
| 1x | 50 | 319.000 | 36634.000 | -19.892 | .000* |
| 1x | 100 | .000 | 36315.000 | -20.069 | .000* |
| 1x | 1000 | .000 | 36315.000 | -20.071 | .000* |
| 2x | 1 | 43863.500 | 88116.500 | -.115 | .908 |
| 2x | 10 | 42994.000 | 87247.000 | -.531 | .595 |
| 2x | 50 | 42123.500 | 86376.500 | -.947 | .344 |
| 2x | 100 | 27318.500 | 71571.500 | -8.026 | .000* |
| 2x | 1000 | .000 | 44253.000 | -21.096 | .000* |
| 5x | 1 | 41866.500 | 86119.500 | -1.070 | .285 |
| 5x | 10 | 33796.500 | 78049.500 | -4.929 | .000* |
| 5x | 50 | 22389.500 | 66642.500 | -10.383 | .000* |
| 5x | 100 | 21090.000 | 65343.000 | -11.005 | .000* |
| 5x | 1000 | 10918.000 | 55171.000 | -15.870 | .000* |

Table: 2 Detailed report of Mann Whitney U test for COR (correlation) at each concentration (1x to 5x) and displacement from 1 to 1000 (in pixels). The significant values are marked with a red colored asterisk.

| Concentration | Displacement (pixels) | Mann-Whitney U | Wilcoxon W | Z | Asymp. Sig. (2-tailed) p value |
|---|---|---|---|---|---|
| 1x | 1 | 34067.000 | 70382.000 | -1.172 | .241 |
| 1x | 10 | 33818.000 | 70133.000 | -1.310 | .190 |
| 1x | 50 | 31579.000 | 67894.000 | -2.552 | .011* |
| 1x | 100 | 31043.000 | 67358.000 | -2.850 | .004* |
| 1x | 1000 | 14081.000 | 50396.000 | -12.258 | .000* |
| 2x | 1 | 42874.000 | 87127.000 | -.588 | .556 |
| 2x | 10 | 41343.000 | 85596.000 | -1.320 | .187 |
| 2x | 50 | 39717.000 | 83970.000 | -2.098 | .036* |
| 2x | 100 | 42385.000 | 86638.000 | -2.822 | .041* |
| 2x | 1000 | 29060.000 | 73313.000 | -7.194 | .000* |
| 5x | 1 | 41857.000 | 86110.000 | -1.075 | .283 |
| 5x | 10 | 43664.000 | 87917.000 | -.211 | .833 |
| 5x | 50 | 41925.500 | 86178.500 | -1.042 | .297 |
| 5x | 100 | 40075.000 | 84328.000 | -1.927 | .054 |
| 5x | 1000 | 27657.000 | 71910.000 | -7.865 | .000* |

Table: 3 Detailed report of Mann Whitney U test for ENT (entropy) at each concentration (1x to 5x) and displacement from 1 to 1000 (in pixels). The significant values are marked with a red colored asterisk.

| Concentration | Displacement (pixels) | Mann-Whitney U | Wilcoxon W | Z | Asymp. Sig. (2-tailed) p value |
|---|---|---|---|---|---|
| 1x | 1 | 31820.000 | 68135.000 | -2.419 | .061 |
| 1x | 10 | 29329.000 | 65644.000 | -3.800 | .000* |
| 1x | 50 | 25264.000 | 61579.000 | -6.055 | .000* |
| 1x | 100 | 23918.000 | 60233.000 | -6.802 | .000* |
| 1x | 1000 | 7871.000 | 44186.000 | -15.703 | .000* |
| 2x | 1 | 41822.000 | 86075.000 | -1.091 | .275 |
| 2x | 10 | 36134.000 | 80387.000 | -3.811 | .000* |
| 2x | 50 | 35133.000 | 79386.000 | -4.290 | .000* |
| 2x | 100 | 42703.000 | 86956.000 | -.670 | .043* |
| 2x | 1000 | 13886.000 | 58139.000 | -14.449 | .000* |
| 5x | 1 | 42020.500 | 86273.500 | -.996 | .319 |
| 5x | 10 | 39893.500 | 84146.500 | -2.014 | .044* |
| 5x | 50 | 34924.000 | 79177.000 | -4.390 | .000* |
| 5x | 100 | 34434.000 | 78687.000 | -4.624 | .000* |
| 5x | 1000 | 26440.000 | 70693.000 | -8.447 | .000* |

Table: 4 Detailed report of Mann Whitney U test for IDM (inverse difference moment) at each concentration (1x to 5x) and displacement from 1 to 1000 (in pixels). The significant values are marked with a red colored asterisk.

Figure 3(I-IV) maps the histograms of the averaged GLCM parameters at various rotations (0° and 90°), displacements (1, 10, 50, 100, and 1000), and initial saline concentrations (Ø = 1x, 2x, and 5x). The error bars represent the standard deviation. These deviations are large as the GLCM parameters change as the droplets dry under ambient conditions. All these parameters show non-significant results (rotationally invariant) when the displacement between the neighboring pixels is 1 at each Ø. The ASM and ENT are only rotationally invariant when the displacement increases to 10 [Fig. 3(I, III)]. These parameters are found to be non-significant when the displacement becomes 50 and 100 only at Ø = 5x. All the GLCM parameters show significant results when the displacement is 1000. The COR at displacement of 1000 pixels in Fig. 3(II) shows negative values for every Ø. It suggests that the relation between the pixels' ceases when the displacement is 1000. Fig. 3(IV) exhibits that the IDM is significantly different for 0° and 90° at all the displacements except 1. This indicates that it is the most sensitive parameter when its values are compared between two different directions. All the parameters except ENT show a decreasing trend in their values when the displacement increases from 1 to 1000. This suggests that the magnitude of the ENT get weakly influenced by the displacement values. Surprisingly, the ASM oriented along 90° at the pixel displacement of 1000 is significantly larger than 0°.

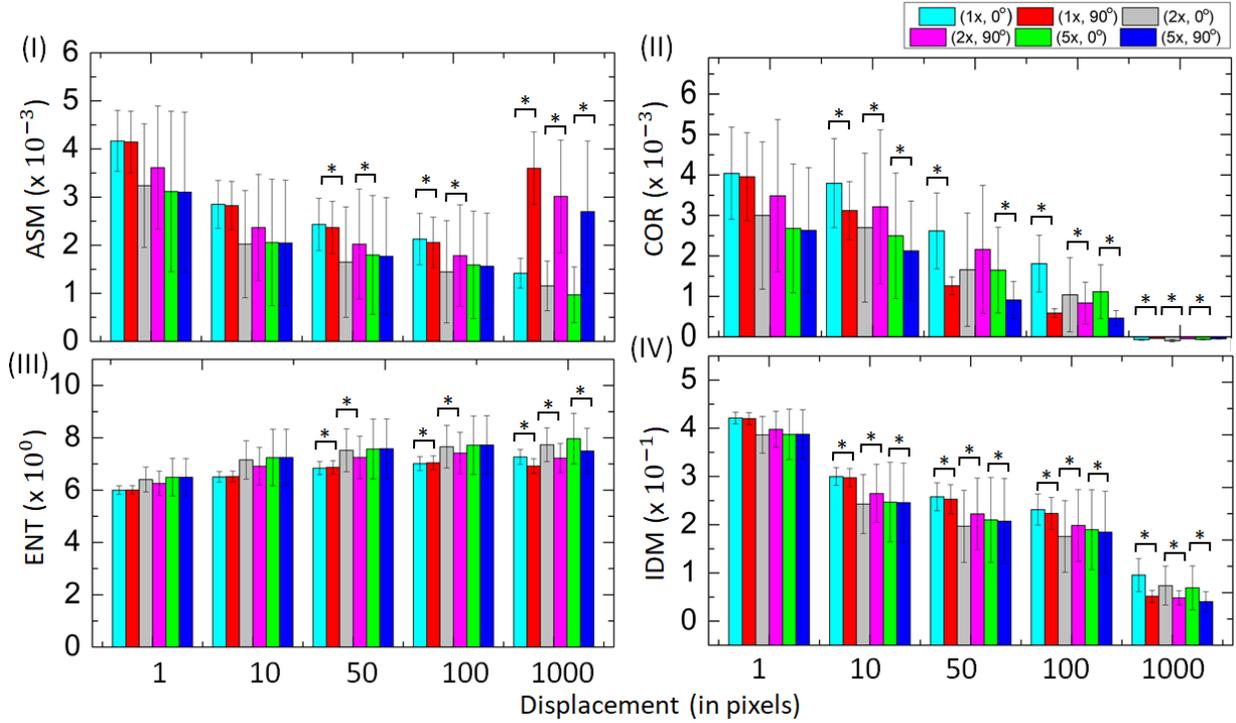

**Figure 3.** (I-IV) show the comparison of the averaged gray level co-occurrence matrix (GLCM) parameters [angular second moment (ASM), correlation (COR), entropy (ENT), and inverse difference moment (IDM)] in arbitrary units (a.u.), respectively in the BSA saline droplets at the different initial saline concentration (Ø) of 1x to 5x. The pixel displacements ranging from 1 to 1000, and the orientations [horizontal (0°) and vertical (90°)] are varied at each Ø. The significant pairs (0° and 90°) are marked with an asterisk (*). The error bars represent the standard deviation.

It can be concluded that these GLCM parameters at 0° and 90° are not significantly different at the displacement of 1 pixel. It suggests that these parameters cannot be averaged for these two orientations except displacement of 1 pixel. Therefore, the averaged GLCM parameters at 1 pixel are presented for the drying evolution at different Ø in Fig. 4(I-IV).

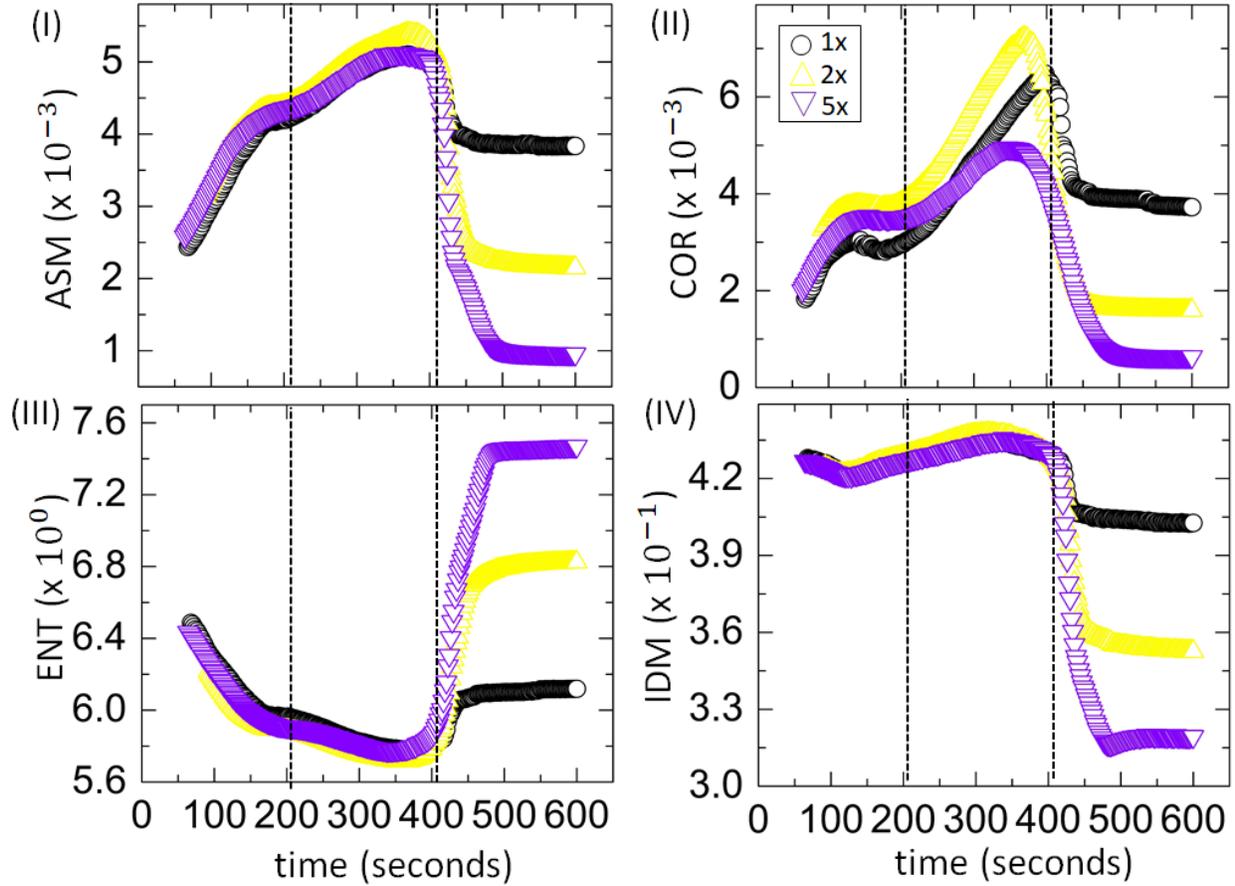

**Figure 4.** (I-IV) shows the time evolution of the gray level co-occurrence matrix (GLCM) parameters [angular second moment (ASM), correlation (COR), entropy (ENT), and inverse difference moment (IDM)] in arbitrary units (a.u.), respectively in the BSA saline droplets as a function of drying time at the different initial saline concentration (Ø) of 1x to 5x. The different stages of the drying process are outlined with the dashed lines in I-IV.

The ASM and the COR increase until ~400 seconds and start decreasing from ~400 seconds onwards [Fig. 4(I-II)]. All the GLCM parameters except COR show a master curve till ~400 seconds. These curves get differentiated after ~400 seconds for the various Ø. On the other hand, the COR shows its dependency throughout the drying process. Interestingly, the ENT [Fig. 4(III)] and the IDM [Fig. 4(IV)] behave differently than the other two parameters. The ENT decreases, whereas the IDM is found to be nearly constant until ~400 seconds. While the ENT starts increasing, the IDM reduces for the duration of ~400 seconds to ~600 seconds. All these parameters, except ENT of the dried films (after the visible drying process ends), show that the values are highest at Ø of 1x, followed by 2x and 5x. Interestingly, the ENT shows a complementary behavior compared to the ASM during the drying process and in the dried states [Fig. 4(I, III)].

## III. DISCUSSIONS

### A. Qualitative analysis of the drying droplets and the dried films

The first image of the drying BSA droplets at different initial saline concentrations (Ø of 1x, 2x, and 5x) is captured soon after their deposition on the substrate (Fig. 1(I-III) a). A higher water loss is observed near the periphery (compared to the central region) due to its hemispherical-cap shape. A capillary flow develops in these droplets. The particles are carried towards the periphery to compensate for this rapid mass loss. This particles relocation continues till the three-phase (solid-liquid, liquid-vapor, and solid-vapor) contact angle reduces significantly. The BSA particles continue to adsorb on the substrate in this initial stage of the drying process. The fluid front recedes from the periphery towards the central region as time advances (Fig. 1(I-III) b), and marks the second stage of the drying process. The two regions, viz., the rim and the central regions become prominent as the front passes and specifies the "coffee-ring effect" [6, 33]. The distinct textures of the regions confirm a greater height of the rim than the central regions. Furthermore, we observed that the width of the rim decreases as the saline concentration rises. Most of the water evaporates from these droplets with the advancement of the drying process. These droplets turn to films, and a mechanical stress develops. Since these droplets are pinned to the substrate (coverslip), these dried films are unable to shrink. The radial cracks originate from the periphery and propagate inwards to relieve the stress from these films (Fig. 1(I-III) c). A few orthoradial cracks also join these radial cracks. This indicates that a dominant stress direction is present in the rim. These events are commonly observed in the BSA droplets (both) in presence and absence of the salts [28, 33]. Hence, it can be argued that the fluid front movement and the developing cracks form the components of the pinned bio-colloidal droplets. We did not observe any evident influence of any salts. The next stage begins with the growing of grainy structures near the dashed line. The grainy structures subsequently cover the entire central region. The dendrite and rosette structures also develop during this stage. The formation of the salt crystals starts towards the later stage of the drying process (Fig. 1(I-III) c-d). This evidently suggests that the drying mechanism is different in the rim and the central regions, and a similar observation is also reported in [28]. The central region is found to be of light gray at Ø of 1x. However, this region turns out to be inhomogeneous, and becomes dark gray with the upsurge of Ø from 2x to 5x. With the increasing numbers of salt crystals, the texture becomes dark gray as the film becomes thick and opaque to pass the light under the transmission mode of the bright-field microscopy. The final morphology is presented at the end of the visible drying process in Fig. 1(I-III) e. It is to be noted that the minute change in the textures is noticed after 24 hours [Figs. 1(I-III) e and 5(a-c)]. The texture of the films becomes darker. This indicates that some water could be trapped in the salt crystals. The evaporation of water after the visible drying process might have caused the textural changes in the dried films.

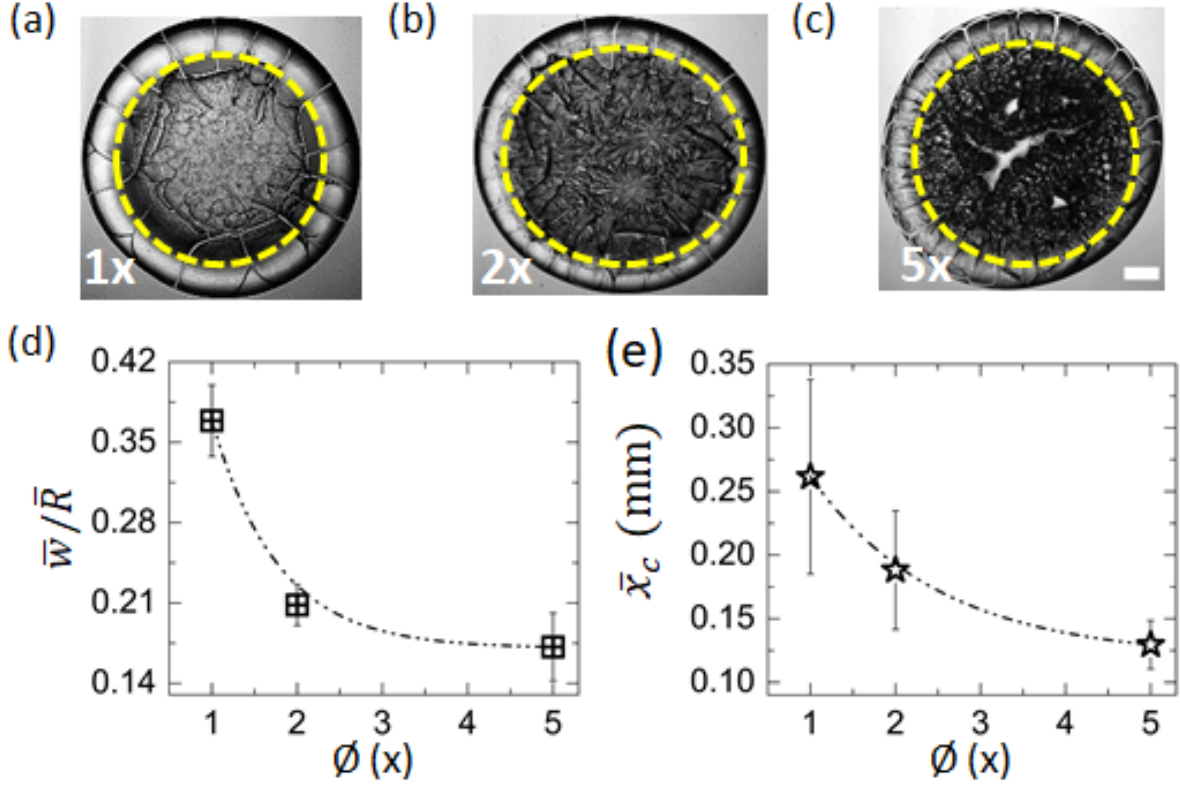

**Figure 5**: The images of the dried films of BSA-saline are captured after 24 hours at various initial concentrations (a) 1x, (b) 2x, and (c) 5x. The yellowed dashed circular lines separate the rim from the central regions. The scale bar has a length of 0.2 mm. The averaged rim width ($\bar{w}$) is divided with the averaged radius of the droplet ($\bar{R}$) and the normalized width ($\bar{w}/\bar{R}$) is shown in (d). The averaged crack spacing ($\bar{x}_c$) is displayed in (e). The error deviations represent the standard deviations. The dashed dotted lines exhibit the exponential decay fitting.

### B. Quantitative analysis of the drying droplets and the dried films

The crack spacing and the ring width are quantified and found to be reduced with the increase of Ø. Fig. 5(d-e) shows the variation of $\bar{w}/\bar{R}$ and $\bar{x}_c$ in the BSA-saline dried films as a function of Ø. Both these parameters decay exponentially with the increment of Ø. The decrease of rim width predominantly takes place due to the reduction of the BSA-BSA interactions in these droplets. As we increase the Ø, the BSA-salt interaction becomes dominant in the formation of the crystal complex structures over the BSA-BSA interactions. Furthermore, the standard deviation of $\bar{x}_c$ is observed to be reduced with the upsurge of Ø. This confirms that the different sized crack domains are present at the low saline concentration (for example, Ø = 1x). Their increased number makes the crack domains uniform when the Ø changes from 1x to 5x. However, it does not mean that the number of cracks increases due to the increase of the saline concentrations. It must be noted that these radial cracks develop only in the rim region, where only the BSA particles are predominantly present. Their number increases because the energy required for the propagation of the smaller cracks is less than the larger cracks.

The FOS parameters (M, SD, SKEW, and KUR) are the spatial moments of order 1 to 4, respectively [32]. These parameters are extracted during the drying evolution of the BSA-saline droplets at Ø of 1x, 2x, and 5x [Fig. 2(I-IV)]. The M is defined as the sum of the gray values of all pixels divided by the number of pixels. It depicts the average values within the selected area (in our case, it is the circular droplet). The SD indicates the amount that the pixel values stray away from the mean. Interestingly, the M shows three ubiquitous stages when the images captured during the drying process are compared (Fig. 1(I-III) a-e). The M grows initially where the contact angle is significantly reduced, and the droplet changes from its spherical-cap shape. In contrast, the uniformity increases, and the SD decreases during this stage. The dark texture of the image changes into bright, and this maps up to ~200 seconds. In the second stage, no change in the M is observed, whereas the SD differentiates the fluid front movement and the origin of the cracks [Fig. 2(I-II)]. It is to be noted that the SD captures most of the textural details even if many complex phenomena assemble them. Both the M and the SD lines up to ~400 seconds that marks the next stage, i.e., the crystal growth. Till this stage, both these parameters are mostly independent of Ø. As the crystals start appearing, the texture transforms (from light) to dark gray. While the M decreases, the complexity increases through the upsurge of the SD.

The SKEW and the KUR of the FOS parameters reveal the information about the distribution of the pixels. The SKEW of an image describes the symmetry of the normal distribution of the pixels in the selected area (in our case, it is the circular droplet). The SKEW of any normal (symmetric) distribution is zero, whereas the negative (positive) values of SKEW indicate that the pixel data are skewed in the left (right) of the distribution. This also implies that the left (right) tail is relatively long than the right (left) tail of its pixel distribution. Comparison of the Figs. 1(I-III) a-e and 2(III) reveal that the distribution of the pixels is asymmetric towards the left throughout the drying process. However, their values change during different stages of drying. In the first stage (within ~200 seconds), the SKEW decreases from -1 a.u. to -4 a.u. The value remains the lowest during the second stage, where the fluid front moves and most of the radial cracks form. With the growth of the crystals in the central region, the SKEW starts increasing, and it reaches to zero only at Ø of 5x. The change in the gaussian distributions from a single to two peaks at gray values of ~15 and ~117 in Fig. 1(III) e is directly connected with the increasing SKEW value towards the end of the drying process. In contrast, the KUR of the image indicates the peak or flatness of the data under consideration. In our case, the KUR shows two peaks at ~150 seconds and ~350 seconds, only to be reduced and reach at a constant value during the drying evolution [Fig. 2(IV)]. These peaks highlight the transition of the first to the second and second to the third drying stages, respectively. More interestingly, the second peak of KUR becomes sharper as the saline concentration (Ø) increases. Since the droplet size is the same for all Ø, the crystals proliferate at Ø = 5x, resulting in a sharp peak during drying.

All these parameters differentiate their values at different Ø as the drying process completes. Nonetheless, the sensitivity of these FOS parameters increases as the higher ranks are computed. This could be a possible reason why only the M displays a uniform distribution, whereas a few peaks and dips are observed in KUR measurements.

The FOS parameters are obtained from the pixel values, it ignores all the tonal and structural properties. In contrast, the tone of the GLCM parameters is based on the pixel intensity (or the gray values). The structure is determined by the spatial relationship between two neighboring pixels. The GLCM parameters, therefore, are based on the local tonal and structural properties that are dependent on each other. The different parameters adopted to extract the FOS and the GLCM properties disable them to compare with each other.

The GLCM parameters (ASM, COR, ENT, and IDM) are extracted during the drying evolution of the BSA-saline droplets at different saline concentrations (Ø of 1x, 2x, and 5x [Fig. 4(I-IV)]). However, it can be concluded from Fig. 3(I-IV) that all the GLCM parameters depend on the orientations and the pixel displacements. Since these are specific to the drying droplets, these parameters need to be checked before making any choice of their computations. The ASM measures the global uniformity of the textures. The uniformity (or the ASM) increases until the salts crystallize in the central region from ~400 seconds. On the other hand, the ENT measures the global heterogeneity of the image. The ASM and ENT are found to be inversely proportional to each other [Fig. 4(I, III)]. The ASM, COR, and IDM parameters show greater uniformity and similarity in the gray regions before the crystallization of the salts occurs. On the other hand, these parameters reduce when the salt crystals start forming due to the appearance of different heterogeneous regions [Fig. 4(I-II, IV)]. This results in decreasing the similarity index between the pixels in a horizontal or vertical orientation. This means that the presence of more salts in the droplets (or the increasing Ø) give rise to more heterogeneity. This leads to the highest value of ENT at Ø = 5x followed by 2x and 1x. Similar arguments also hold for other parameters (ASM, COR, and IDM). The Ø of 5x shows the lowest values than 2x or 1x when ASM, COR, and IDM are computed towards the end of the drying process.

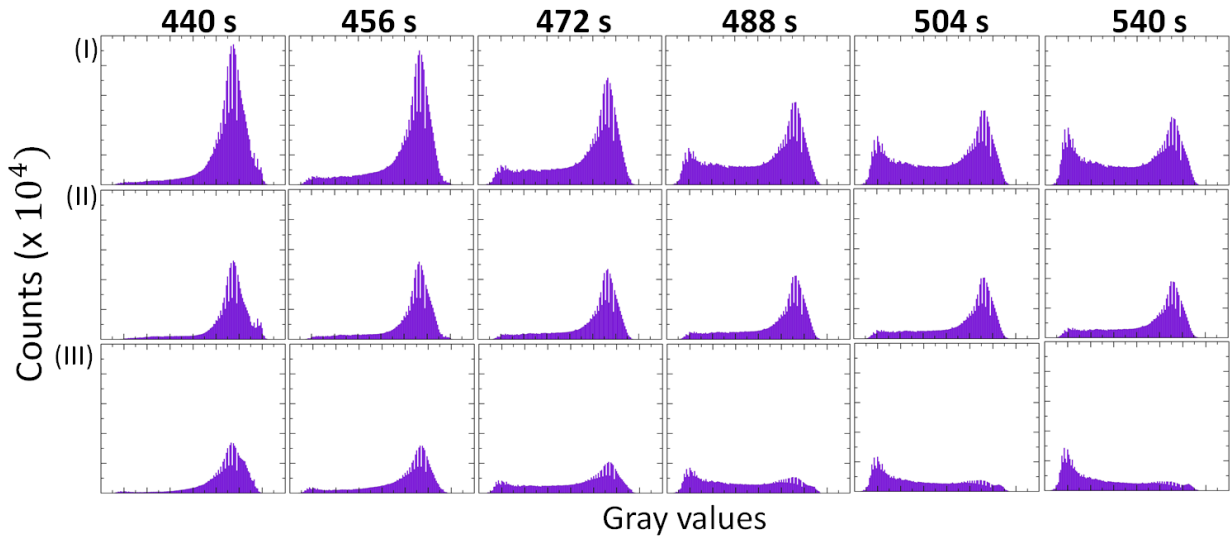

**Figure 6:** Histograms depicting the counts of the pixels along the y-axis, and the gray values along the x-axis of each image taken at 440 s, 456 s, 472 s, 488 s, 504 s, and 540 s at Ø = 5x for (I) whole droplet, (II) rim region, and (III) central region. The gray values in x-axis ranges from 0 to 160 in the interval of 20. The pixel counts in y-axis ranges from 0 to 25 (x $10^4$) in the interval of 5 (x $10^4$).

To understand the physical mechanism of the appearance of such a bi-modal distribution at Ø = 5x, we plotted its progression from 440 seconds to 540 seconds [Fig. 6(I-III)]. It is clear from Fig. 1(III) d-f that these two peaks origin during the crystal growth stage, somewhat at the very end of the drying process. This is the stage when almost all the water evaporates from these droplets. The salt layer is formed in the central region during this time. More crystals-like structures appear as the trapped water evaporates from these droplets. It means that the equilibrium-like regions start emerging locally, whereas the whole system (drying droplet) is still out of the equilibrium. Two local equilibrium-like regions (one in the rim, and the other in the central region) appear when the droplets approach the steady-state (end of the drying process). However, these two regions seem to be weakly correlated. To validate this assumption, we extracted the pixel counts as a function of the gray values only for the rim [Fig. 6(II)] and the central regions [Fig. 6(III)]. It is observed that the second peak decreases to the minimum and the first peak appears at 472 seconds when the counts are computed for the central region. In contrast, the rim does not show any such behavior and consistently show the second peak. This implies that the gray value of ~15 (first peak) is due to the emergence of the crystal structures in the central region. The second peak at gray value of ~117 confirms the origin of the rim region. If these regions were uncorrelated, zero values in their interaction terms should have emerged. However, we observed a finite value between the peaks when the pixel counts are plotted for the whole droplet [Fig. 6(I)]. This confirms that the BSA-BSA interactions are dominant (recessive) over the BSA-saline interactions in the rim (central) regions that results in the phase separated aggregation of the BSA particles in presence of the salts. It is also understood that the saline concentration, in principle, should have been enough for the BSA-saline interactions to be dominant over the BSA-BSA interactions. This is probably the reason that we did not observe any bimodal distributions in the lower saline concentrations. Furthermore, it also validates why Ø of 5x behaves uniquely when the FOS and GLCM parameters are extracted during the drying evolution.

## IV. CONCLUSIONS

This chapter reveals that the image processing techniques can be used to understand and quantify the textural features emerged during the drying process. This image processing methodology adopted in this chapter is certainly useful in quantifying the textural changes of the patterns at different saline concentrations that dictates the ubiquitous stages of the drying process. The first-order statistical parameters such as the standard deviation captures the textural complexity, whereas the *mean* of the image averages the textural details. Contrary to the first order statistical analysis, the gray level co-occurrence matrix (GLCM) summarizes both the tonal and structural relationships between the neighboring pixels. The horizontal and the vertical orientations of the GLCM parameters show a non-significant effect when the pixel displacement is $\leq 1$. The entropy and the angular second moment of GLCM appeared to be inversely proportional during the evolution process. The distribution changes from single to bimodal as the initial saline concentration increases. This helps us to understand the aggregation process of the BSA particles in the presence of the salts. This chapter, thus, demonstrates the versatility and usefulness of these image processing techniques, and it assumes that similar methodology can be adopted in other (related) bio-colloidal systems to set-up a successful clinical settings in the near future.

**ACKNOWLEDGMENTS**

This work is supported by the Department of Physics at WPI. The authors are grateful to the grant received from the WPI Tinkerbox community sponsored by the WIN (Women Impact Network).